\documentclass[11pt]{article}
\usepackage{graphicx}
\usepackage{amssymb}
\usepackage{epstopdf}
\title{From ``Not Wrong" to (Maybe) Right\footnote{This is, in essence, a solicited ``Turning Points" feature written for Nature. It appeared, in a slightly abbreviated form, in the March 18 issue.}}
\author{Frank Wilczek\\
{\small\itshape Center for Theoretical Physics}\\
{\small\itshape Department of Physics}\\
{\small\itshape Massachusetts Institute of Technology}\\
{\small\itshape Cambridge, MA 02139-4307}\\
{\small\itshape MIT-CTP-3485}}
\date{}
\begin{document}
\maketitle
\begin{abstract}
This is a short, light spirited account of how some possibly important science actually happened. It very much conflicts with Popper's contention that the key to scientific progress is falsification.
\end{abstract}
Savas Dimopoulos is always enthusiastic about something, and in the spring of 1981 it was supersymmetry. He was visiting the new Institute for Theoretical Physics in Santa Barbara, which I had recently joined.  We hit it off immediately -- he was bursting with wild ideas, and I liked to stretch my mind by trying to take them seriously. 

Supersymmetry was (and is) a beautiful mathematical idea.  It extends the symmetry of special relativity.  Special relativity postulates the invariance of physical law under motion with a constant velocity. It thereby allows us to transform between objects with different speeds and to predict the properties of moving particles from those of stationary ones. Supersymmetry postulates the invariance of physical law under certain kinds of motion in a quantum-mechanical extension of space-time, superspace.  Superspace has four extra quantum-mechanical dimensions, quite different from the familiar four dimensions of space and time. When a particle ``moves" in the extra quantum dimensions it just acquires a tiny amount of angular momentum, or spin. So you wouldn't want to live in the new suburbs of superspace: it's very cramped, and makes you dizzy. But the mathematics of supersymmetry promised (and still promises) to help us unify fundamental physics: it allows us to transform between objects of different spin. Several different kinds of particles, differing in their spin, could be manifestations of a single entity moving at different speeds through superspace.

The problem with applying supersymmetry is that it's too good for this world. We simply don't find particles of the sort it predicts. We don't, for example, see particles with the same charge and mass as electrons, but a different amount of spin.   

Symmetry principles that might help to unify fundamental physics are hard to come by, however, so theoretical physicists won't give up on them easily. Based on previous experience with other forms of symmetry, we've developed a fallback strategy, called spontaneous symmetry breaking.  In this approach we postulate that the fundamental equations of physics have the symmetry, but the stable solutions of these equations do not.   The classic example of this phenomenon occurs in an ordinary magnet: in the basic equations that describe the physics of a lump of iron any direction is equivalent to any other, but the lump becomes a magnet with some definite north-seeking pole.  

While straight supersymmetry is a (wrong) statement about the properties of the world, spontaneously broken supersymmetry is a research program. For it is a statement about equations that describe the world, and only indirectly about the world itself. Carrying forward this research program involves model
building -- the creative activity of proposing candidate equations and analyzing their consequences. 

Building models with spontaneously broken supersymmetry that are consistent with everything else we know about physics is a difficult business. Even if you manage to get the symmetry to break, the extra particles are still there (just heavier) and cause various mischief.  I briefly tried my hand at model-building when supersymmetry was first developed, mainly by Julius Wess and Bruno Zumino \cite{fw1} in the mid-1970s, but after some simple attempts failed miserably I gave up.  

Savas was a much more naturally gifted model-builder, in two crucial respects: he didn't insist on simplicity, and he didn't give up.  When I identified some difficulty (let's call it A) that was not addressed in his model, he would say ``It's not a real problem, I'm sure I can solve it" and the next afternoon he would come in with a more elaborate model that solved difficulty A. But then we'd discuss difficulty B, and he'd solve that one with a completely different complicated model. Now to solve both A and B, you had to join the two models, and so on to difficulty C, and soon things got incredibly complicated.  Working through the details, we'd find some flaw.  Then the next day Savas would come in, very excited and happy, with an even more complicated model that fixed yesterday's flaw. Eventually we eliminated all flaws, using the method of proof by exhaustion: anyone, including us, who tried to analyze the model would get exhausted before they understood it well enough to find the flaws.

When I tried to write up what we'd done for publication, there was a certain feeling of unreality and embarrassment about the complexity and arbitrariness of what we'd come up with. I couldn't help but compare our mess to some truly elegant ideas about unification, which involved more conventional kinds of gauge symmetry, not supersymmetry \cite{fw2, fw3}.  Those ideas were genuinely fruitful, in that they explained some things we already knew (the relative values of the strong, electromagnetic, and weak coupling constants) \cite{fw4}.   

Savas was undaunted. He maintained that those ideas weren't really so elegant, if you tried to be completely realistic and work them out in detail. He'd been talking to another colleague, Stuart Raby, about trying to improve those models by adding supersymmetry! I was extremely skeptical about this ``improvement", because I was certain that the added complexity of supersymmetry would spoil the existing success of gauge symmetry in explaining the couplings.  The three of us decided to do the calculation, to see how bad the situation was. To get oriented and make a definite calculation, we started by doing the crudest thing, that is to ignore the whole problem of breaking supersymmetry, which allowed us to use very simple (but manifestly unrealistic) models.   

The result was amazing, at least to me. The supersymmetric versions of the gauge symmetry models, although they were vastly different from the originals, gave very nearly the same answer for the couplings. 

That was the turning point. We put aside the ``not wrong" complicated models with spontaneous supersymmetry breaking, and wrote a short paper \cite{fw5} that, taken literally (with unbroken supersymmetry), was wrong. But it presented a result that was so straightforward and successful that it made the idea of putting gauge symmetry and supersymmetry unification together seem (maybe) right. We put off the problem of how supersymmetry gets broken. And while there are some good ideas about it, there is still no generally accepted solution. 

Subsequently, more precise measurements of the couplings made it possible to distinguish between the predictions of models with and without supersymmetry \cite{fw6}.   The models with supersymmetry work much better.  We all eagerly await operation of the Large Hadron Collider (LHC) at CERN, where, if these ideas are correct, the new particles of supersymmetry -- or, you might say, the new dimensions of superspace -- must make their appearance.  

This little episode, it seems to me, is 179 degrees or so out of phase from Popper's idea that we make progress by falsifying theories. Rather in many cases, including some of the most important, we suddenly decide our theories might be true, by realizing that we should strategically ignore glaring problems. It was a similar turning point when David Gross and I decided to propose QCD based on asymptotic freedom, putting off the problem of quark confinement. But that's another story ...

 \end{document}